\documentclass[aps, prl, superscriptaddress, twocolumn]{revtex4-1}
\usepackage[latin9]{inputenc}
\usepackage[letterpaper]{geometry}
\usepackage{color}
\usepackage{amsmath}
\usepackage{amssymb}
\usepackage{graphicx}
\usepackage[unicode=true,pdfusetitle,
 bookmarks=true,bookmarksnumbered=false,bookmarksopen=false,
 breaklinks=false,pdfborder={0 0 1},backref=false,colorlinks=true]
 {hyperref}
\hypersetup{
 pdfborderstyle=,linkcolor=blue,urlcolor=blue,citecolor=blue}

\makeatletter
\usepackage{float}
\usepackage{upgreek}
\usepackage{braket}
\usepackage{upgreek}
\usepackage{lipsum}

\makeatother

\begin{document}
\title{Unusual Diffusivity in Strongly Disordered Quantum Lattices: Random
Dimer Model}
\author{Ilia Tutunnikov}
\affiliation{Department of Chemistry, Massachusetts Institute of Technology, 77
Massachusetts Avenue, Cambridge, Massachusetts 02139, USA \looseness=-1}
\affiliation{ITAMP, Center for Astrophysics \textbar{} Harvard \& Smithsonian,
Cambridge, Massachusetts 02138, USA\looseness=-1}
\author{Jianshu Cao}
\email{jianshu@mit.edu}

\affiliation{Department of Chemistry, Massachusetts Institute of Technology, 77
Massachusetts Avenue, Cambridge, Massachusetts 02139, USA \looseness=-1}
\begin{abstract}
Recent advances in transport properties measurements of disordered
materials and lattice simulations, using superconducting qubits, have
rekindled interest in Anderson localization, motivating our study
of highly disordered quantum lattices. Initially, our statistical
analysis of localized eigenstates reveals a distinct transition between
weak and strong disorder regimes, suggesting a random distribution
of dimers in highly disordered systems. Subsequently, the random dimer
model predicts an oscillating diffusivity that decays as $t^{-1/2}$,
is inversely proportional to the disorder strength, and maintains
a constant frequency with an initial phase shift of $\pi/4$. The
first peak exhibits a universal scaling of $\sigma^{-1}$ both in
peak time and amplitude. Finally, we find that stochastic noise suppresses
these oscillations and induces hopping between localized eigenstates,
resulting in constant diffusion over long times. Our predictions challenge
the conventional understanding of incoherent hopping under strong
disorder. This offers new insights to optimize disordered systems
for optoelectrical and quantum information technologies.
\end{abstract}
\maketitle
Nature abounds in various forms of disorder, evident in both natural
and synthetic materials. In his seminal work \citep{Anderson1958},
P. W. Anderson introduced the concept of quantum (Anderson) localization
of a single electron wave function in a disordered lattice. Experimentally,
observing localization remains challenging due to interparticle interactions
and lattice fluctuations \citep{Abanin2019}. Despite these challenges,
localization has been demonstrated experimentally in various systems
ranging from Bose-Einstein condensates \citep{Roati2008,Billy2008}
to disordered quantum circuits \citep{Karamlou2022}. In a broader
context, localization is now recognized as a universal wave phenomenon
that can be realized with microwaves \citep{Chabanov2000}, ultrasound
waves \citep{Hu2008}, and light \citep{Segev2013}, and it plays
a pivotal role in the optical and transport properties of molecular
solids \citep{DavydovBook1971,Silbey1976,KuhnBook2011,NitzanBook2006}.
Theoretically, our understanding has advanced significantly through
the development of scaling theory \citep{Abrahams1979} and quantum
phase transition theory \citep{Evers2008}, among others \citep{50YearsBook}.
Crucially, in one-dimensional (1D) systems, any amount of disorder
can localize all eigenstates, whereas in 3D, localization depends
critically on disorder strength. 

While numerous studies have focused on weak disorder \citep{Yamada1992,Skipetrov2004,Cui2023}
and diffusive \citep{Aslangul1989,Cao2009,Hoyer2010,Moix2012,Chuang2016,Cao2020}
regimes, the transient diffusivity in highly-disordered 1D quantum
lattices remains poorly understood, which defines the focus of this
Letter. Our study reveals the nature of the oscillating diffusivity,
which differs markedly from diffusivity in ordered or weakly disordered
lattices. Furthermore, thermal fluctuations suppress the oscillations
in diffusivity and induce hopping between localized eigenstates, which
leads to constant diffusion in the long-time limit. These diverse
phenomena are unified within the proposed `random dimer model', which
challenges the conventional picture of incoherent hopping in the strong
disorder regime and aids in interpreting recent measurements in quantum
dot superlattices \citep{Blach2022,Blach2023} and superconducting
circuits \citep{Karamlou2022}.\\

\emph{Stationary distribution in a closed system.}--We consider an
isolated 1D quantum lattice consisting of $N$ sites described by
the Anderson Hamiltonian \citep{Anderson1958,Thouless1974}
\begin{align}
\hat{H} & =\sum_{n}\epsilon_{n}\ket{n}\bra{n}+J\sum_{n}(\ket{n}\bra{n+1}+\ket{n+1}\bra{n}),\label{eq:hamiltonian}
\end{align}
where the site energies $\epsilon_{n}$ are Gaussian random variables
with zero mean and standard deviation $\sigma$, and $J$ is the coupling
constant. In this Letter, we set $\hbar=1$, energy is measured in
units of $|J|$, and time is measured in units of $1/|J|$. Additionally,
the lattice is equally spaced with lattice constant $a$, and distances
are measured in units of $a$.

\begin{figure}[t]
\begin{centering}
\includegraphics{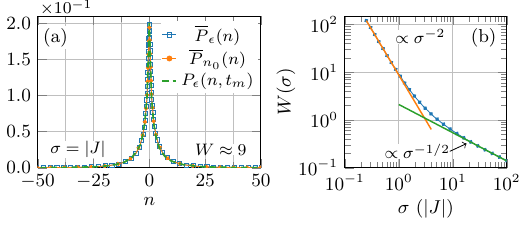}
\par\end{centering}
\caption{(a) Average stationary population distributions. $P_{\epsilon}(n,t_{m})$
is added for reference, and is obtained by direct solution of the
time-dependent Schr\"odinger equation and averaging over disorder
realizations; $t_{m}$ is sufficiently large time. Here, $N=201$,
$\sigma=|J|$. (b) Standard deviation of the average stationary distribution
as a function of disorder strength. \label{fig:FIG1}}
\end{figure}

When the 1D system is initiated with a single populated site $n=n_{0}$
at $t=0$, a finite size stationary state is eventually reached for
any disorder strength, $\sigma>0$ \citep{Abrahams1979,DominguezAdame2004}.
The population distribution can be characterized in two approximately
equivalent ways. First, we carry-out time averaging
\begin{align}
\overline{P}(n-n_{0};n_{0}) & =\lim_{T\rightarrow\infty}\frac{1}{T}\int_{0}^{T}P(n-n_{0},t;n_{0})\,dt\nonumber \\
 & =\sum_{k=1}^{N}|\braket{n_{0}|v_{k}}|^{2}|\braket{n-n_{0}|v_{k}}|^{2},\label{eq:sites-pop-distribution-time-averaged}
\end{align}
where $\ket{v_{k}}$ are the eigenstates of $\hat{H}$. Then, we consider
two average distirbutions: (i) $\overline{P}(n;n_{0}=0)$ averaged
over the random site energies, $\overline{P}_{\epsilon}$, and (ii)
$\overline{P}(n-n_{0};n_{0})$ averaged over the initial position
$n_{0}$, $\overline{P}_{n_{0}}(n)$. $\overline{P}_{\epsilon}$ and
$\overline{P}_{n_{0}}$ are approximately equal {[}see Fig. \ref{fig:FIG1}(a){]},
because moving the initial site $n_{0}$ is equivalent to generating
a new random sequence of site energies. In the thermodynamic limit
$N\rightarrow\infty$, the two distributions become equal, consistent
with the concept of self-averaging.

There are several metrics for characterization of the stationary state
width \citep{Moix2012,Moro2018}. Here, we use the standard deviation
(with $\overline{P}_{\epsilon}$, for convenience)
\begin{equation}
W(\sigma)=\sqrt{\sum_{n}n^{2}\overline{P}_{\epsilon}},\label{eq:mean-squared-displacement}
\end{equation}
Figure \ref{fig:FIG1}(b) suggests a qualitative transition between
weak and strong disorder regimes. For relatively weak disorder ($\sigma<|J|$),
$W$ scales approximately as $\sigma^{-2}$ (localization length scaling
in 1D quantum lattice \citep{Moix2013}) . For strong disorder ($\sigma>10|J|$),
the distributions is highly localized ($W<1$), and $W$ scales approximately
as $\sigma^{-1/2}$. See Supplemental Material (SM) \citep{SuppMat}
for further qualitative discussion of the scaling.

\begin{figure}[t]
\begin{centering}
\includegraphics{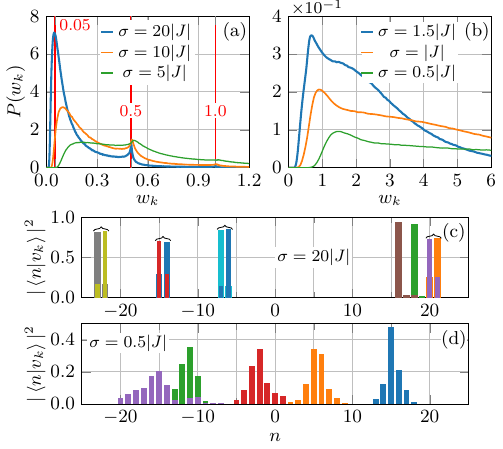}
\par\end{centering}
\caption{(a, b) Distributions of eigenstate standard deviations, $w_{k}$ {[}see
Eq. \eqref{eq:eigenstates-widths}{]} for several values of disorder
strength, $\sigma$. (c) Ten eigenstates with $w_{k}\approx0.5$ in
a lattice of $N=51$ sites at $\sigma=20|J|$. (d) Five eigenstates
in a lattice of $N=51$ sites at $\sigma=0.5|J|$. \label{fig:FIG2}}
\end{figure}

A more detailed picture is revealed when considering the standard
deviations of eigenstates $\ket{v_{k}}$, defined as
\begin{equation}
w_{k}^{2}=\sum_{n}n^{2}|\!\braket{n|v_{k}}\!|^{2}-\left[\sum_{n}n|\!\braket{n|v_{k}}\!|^{2}\right]^{2}.\label{eq:eigenstates-widths}
\end{equation}
Figure \ref{fig:FIG2}(a) shows that for $\sigma=20|J|$ the distribution
of $w_{k}$ has a major peak at $w_{k}\approx0.05$ and a minor peak
at $w_{k}\approx0.5$, and it rapidly decays for $w_{k}>0.5$. The
major peak at $w_{k}=0.05$ corresponds to states consisting of practically
a single site, i.e., monomers. Figure \ref{fig:FIG2}(c) shows ten
eigenstates with $w_{k}\approx0.5$ that mostly belong to dimers denoted
by curly brackets. Evolution of the initial state can be described
by representing the initial state as a superposition of lattice eigenstates.
Due to the demonstrated effective partitioning of the lattice into
monomers and dimers, the wave packet size is limited to a few sites.

As $\sigma$ decreases, the fraction of monomers diminishes. The tail
of the distribution becomes thicker for $\sigma<10|J|$, and the dimer
picture breaks down. Figure \ref{fig:FIG2}(b) shows the distributions
for lower disorder strength. In this regime, the distributions have
a single peak which shifts to the right with decreasing $\sigma$.
Figure \ref{fig:FIG2}(d) shows five eigenstates at $\sigma=0.5|J|$
spanning multiple sites.\\

\emph{Transient dynamics in the strong-disorder regime.}-- To analyze
the transient wave packet spreading, we focus on the time-dependent
rate of wave packet expansion, i.e., the diffusivity
\begin{equation}
2D(t)=\frac{d\braket{n^{2}}}{dt}-\frac{d\braket{n}^{2}}{dt},\label{eq:D(t)-definition}
\end{equation}
where $\braket{n^{k}}=\sum_{n}n^{k}P(n,t)$. In an ordered system,
a localized initial state expands ballistically, i.e., $\braket{n^{2}}=2J^{2}t^{2}$,
and $D(t)=2J^{2}t$. The ballistic expansion results from the overlap
between the initial state with all the extended eigenstates of the
lattice (Bloch states). In contrast, in a disordered 1D system, all
the eigenstates have finite extent \citep{Anderson1958,Thouless1974,DominguezAdame2004}
resulting in transient expansion and eventual localization.

As suggested earlier, for sufficiently strong energy disorder, the
chain is effectively partitioned into monomers and dimers. Thus, we
can approximate the diffusivity in the chain by the diffusivity of
a random dimer \citep{Mazza1998}. This model traces its root to the
standard tunneling model, which was originally proposed to predict
the thermal and transport properties of amorphous solids \citep{Phillips1972,P.W.Anderson1972,Heuer1993}.
Most significant is the application of the model to explain the unusual
spectral diffusion measured by early single molecule experiments \citep{Geva1996,Moerner1999}.
Evidently, the dimers in our highly disordered Anderson model give
rise to the random distribution of the incoherent two-level systems
(TLS) in the standard tunneling model. 

Let the two sites be labeled by $0$ and $1$, and the initial state
$P(n,t=0)=\delta_{n0}$, the population on site $n=1$ at time $t$
reads
\begin{equation}
P_{1}(t)=\frac{2J^{2}}{4J^{2}+\epsilon_{01}^{2}}\left[1-\cos\left(t\sqrt{4J^{2}+\epsilon_{01}^{2}}\right)\right],
\end{equation}
where $\epsilon_{01}=\epsilon_{0}-\epsilon_{1}$. Note that the monomers
are correctly accounted for in this model -- when the energy difference
between the two sites is much higher than $|J|$, the bond is effectively
broken. The disorder-averaged diffusivity reads
\begin{align}
D(t) & =\frac{1}{2\sigma\sqrt{\pi}}\int_{-\infty}^{\infty}\dot{P}_{1}(t)\exp\left(-\frac{\epsilon_{01}^{2}}{4\sigma^{2}}\right)\,d\epsilon_{01},\label{eq:dimer-D(t)-disorder-ave}
\end{align}
where we used the fact that $\epsilon_{01}$ is a normal random variable
with variance $2\sigma^{2}$. Asymptotically (for sufficiently large
$t$ and $\sigma\gg|J|$, see SM \citep{SuppMat} for details),
\begin{equation}
D_{A}(t)\sim\frac{|J|^{2/3}}{\sigma}\frac{\sin(2|J|t+\pi/4)}{\sqrt{t}}.\label{eq:D_A(t)}
\end{equation}

\begin{figure}[h]
\begin{centering}
\includegraphics{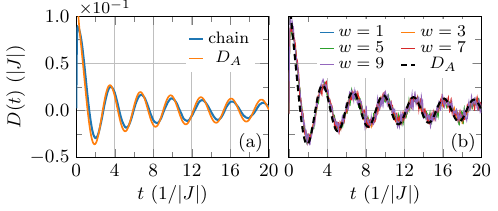}
\par\end{centering}
\caption{(a) Diffusivity of a localized initial state in a lattice of $N=21$
sites vs the asymptotic diffusivity of a disordered dimer, $D_{A}(t)$
{[}see Eq. \eqref{eq:D_A(t)}{]}. (b) Diffusivity of Gaussian initial
state of width $w$ vs $D_{A}(t)$. Superimposed high frequency noise
diminishes with increasing number of site energies realizations used
in averaging. Here, $\sigma=20|J|$. \label{fig:FIG3}}
\end{figure}

Figure \ref{fig:FIG3}(a) compares the diffusivity obtained numerically
by solving the time-dependent Schr\"odinger equation for a long lattice
vs the asymptotic formula in Eq. \eqref{eq:D_A(t)}. For $t>1/|J|$,
the diffusivity oscillates at frequency $2|J|$; the amplitude is
proportional to $(\sigma\sqrt{t})^{-1}$; and the first peak introduces
a phase shift of $\pi/4$. This behaviour differs markedly from the
behaviour at low-moderate disorder, where the diffusivity has a single
peak, and decays steadily to zero after that \citep{Blach2023,CaoGroup}.
The discrepancy between the numerically exact diffusivity and Eq.
\eqref{eq:D_A(t)} grows with decreasing $\sigma$, due to the increasing
contributions from larger clusters (trimers, etc., see SM for an example
\citep{SuppMat}). Within the random dimer model, $W\propto\sqrt{D_{A}}\propto\sigma^{-1/2}$
consistent with the scaling in Fig. \ref{fig:FIG1} (also see SM \citep{SuppMat}).

\begin{figure}[h]
\begin{centering}
\includegraphics{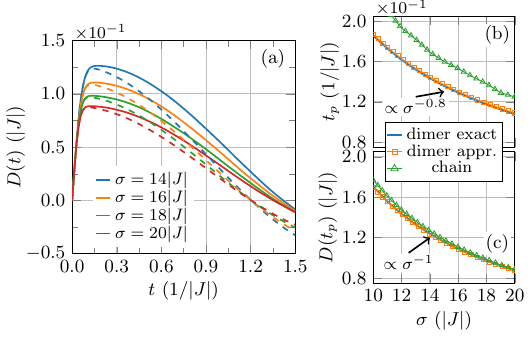}
\par\end{centering}
\caption{(a) Diffusivity near turnover, numerical results in a chain (solid)
vs dimer model (dashed). (b, c) Numerically exact results in dimer
(solid) and chain (triangles) vs. Eqs. \eqref{eq:D_S(t)} and \eqref{eq:t_p}
(squares).  \label{fig:FIG4}}
\end{figure}

Close to $t=0$, the diffusivity grows linearly, and $\dot{D}=2J^{2}$
as in an ordered lattice. After the first peak, the diffusivity continues
approximately according to Eq. \eqref{eq:D_A(t)}. In the turnover
region, the diffusivity is approximately given by (see SM for details
\citep{SuppMat})
\begin{align}
D(t) & \approx\frac{\sqrt{\pi}J^{2}}{\sigma}(1-J^{2}t^{2})\text{erf}(\sigma t).\label{eq:D_S(t)}
\end{align}
Figure \ref{fig:FIG4}(a) compares $D(t)$ near the turnover in a
chain and a dimer. For $\sigma/|J|\geq15$, the approximate expression
captures the peak height relatively well. The turnover time, $t_{p}$
(a point where $\dot{D}=0$) is approximately given by (see SM for
details \citep{SuppMat})
\begin{equation}
t_{p}\approx\frac{1}{\sigma\sqrt{2}}\sqrt{\ln\left(\frac{2\sigma^{4}}{\pi J^{4}}\right)-\ln\left[\ln\left(\frac{2\sigma^{4}}{\pi J^{4}}\right)\right]}.\label{eq:t_p}
\end{equation}
With increasing $\sigma$, $t_{p}$ scales slower than $\sigma^{-1}$.
Figures \ref{fig:FIG4}(b) and \ref{fig:FIG4}(c) compare the numerically
exact $t_{p}(\sigma)$ and $D(t_{p})$ in a dimer and a chain vs the
analytical approximations. While the dimer model captures the decay
of $t_{p}(\sigma)$ correctly, the values are slightly underestimated,
since the chain still contains some fraction of larger clusters (trimers,
etc.).

Preparing an initial state with a single excited site is feasible
in special systems, e.g., in highly controllable quantum superconducting
circuits \citep{Karamlou2022}. In contrast, optical excitation results
in initial spatially extended states \citep{Blach2022,Blach2023}.
Generally, in presence of either disorder or noise, the transient
diffusivity of spatially extended initial states qualitatively differs
from the diffusivity of a localized initial state \citep{Cui2023,Tutunnikov2023}.
In the strong disorder regime, however, the picture simplifies considerably.
Figure \ref{fig:FIG3}(b) shows the time-dependent diffusivity of
Gaussian initial states of various initial widths, $w$. Diffusivity
is nearly independent of $w$, which stems from the effective partitioning
of the lattice into monomers and dimers due to disorder. The extended
initial state overlaps with multiple localized eigenstates, and the
disorder averaging results in a diffusivity close to the diffusivity
of a random dimer, $D_{A}(t)$ in Eq. \eqref{eq:D_A(t)}. \\

\emph{Diffusivity in an open system.}-- Next, we consider the diffusivity
in an open quantum system coupled to a noisy environment. Long-time
diffusion in strongly disordered lattices is an intriguing problem,
where traditional random walk models, such as the Forst or Marcus
rate, fail to capture transient quantum coherence features, like slow
non-exponential relaxation, and quantum beatings. Here, the environment
influence is accounted for by introducing stochastic site energy fluctuations
within the framework of the Haken-Strobl-Reineker (HSR) model \citep{ExcitonDynamicsBook1982}.
In the HSR model the reduced density matrix obeys the Lindblad master
equation
\begin{align}
\dot{\rho} & =-i[H,\rho]-\frac{\Gamma}{2}\sum_{n}[V_{n},[V_{n},\rho]],\label{eq:HSR-model}
\end{align}
where $(V_{n})_{j,k}=\delta_{j,n}\delta_{k,n}$, and $\Gamma$ is
the dephasing rate. We can rewrite Eq. \eqref{eq:HSR-model} as $\dot{\rho}=-i[H,\rho]-\Gamma\rho_{H},$where
$\rho_{H}$ is a `hollow density matrix', with elements $(\rho_{H})_{m,n}=(1-\delta_{m,n})\rho_{m,n}.$
The latter form shows that the $\Gamma$ term damps the off-diagonal
elements of the density matrix (coherences). The diffusivity of a
localized initial state in the ordered HSR model is given by \citep{Moix2013}
$D(t)=(2J^{2}/\Gamma)[1-\exp(-\Gamma t)].$ On the short time scale
($\Gamma t\ll1$), when the dephasing effect is negligible, one recovers
the ballistic behavior, i.e., $D(t)\propto t$. In contrast, on the
long time scale, the coherence is lost and $D(t\rightarrow\infty)=2J^{2}/\Gamma$.

\begin{figure}[h]
\begin{centering}
\includegraphics{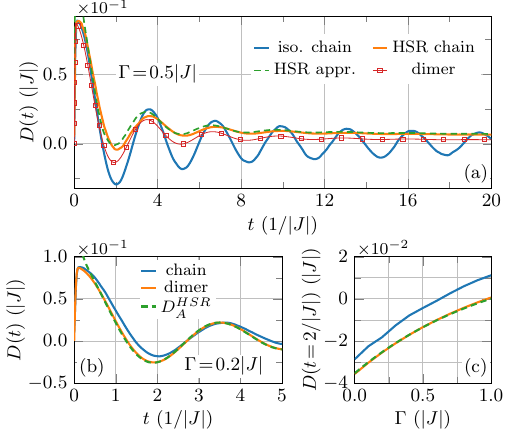}
\par\end{centering}
\caption{(a) Diffusivities in isolated and HSR lattices ($N=21$). (b) Diffusivity
in disordered HSR lattice compared to $D_{A}^{HSR}$ in Eq. \eqref{eq:D_A_HSR(t)}.
(c) First minimum of $D(t)$ as a function of $\Gamma$. Here, $\sigma=20|J|$.
\label{fig:FIG5}}
\end{figure}

Figure \ref{fig:FIG5}(a) compares the diffusivities in isolated and
HSR lattices in the regime of strong disorder. In contrast to the
closed system, the diffusivity oscillations are further damped, and
it slowly tends to a non-zero steady-state value. Asymptotically,
the diffusivity of a random HSR dimer is a sum of two terms (see SM
for details \citep{SuppMat})
\begin{subequations}
\begin{align}
D_{A}^{HSR}(t) & =D_{1}^{HSR}(t)+2\mathrm{Re}[D_{2}^{HSR}(t)],\label{eq:D_A_HSR(t)}\\
D_{1}^{HSR}(t) & \sim\frac{|J|}{2\sigma}\sqrt{\frac{\Gamma}{t}}-\frac{J^{2}\Gamma}{\sigma^{2}},\label{eq:D1_HSR(t)}\\
2\mathrm{Re}[D_{2}^{HSR}(t)] & \sim\frac{\sqrt{2}J^{2}e^{-\Gamma t/2}}{\sigma\sqrt{t}(\Gamma^{2}+4J^{2})^{1/4}}\nonumber \\
 & \times\sin\left[2t|J|\!-\!\frac{1}{2}\arg(\Gamma\!-\!2i|J|)\right],\label{eq:D2_HSR(t)}
\end{align}
\end{subequations}
where $D_{1}^{HSR}(t)$ is a slowly decaying baseline, and $D_{2}^{HSR}(t)$
describes the oscillations. Figure \ref{fig:FIG5}(b) compares the
exact diffusivity of a HSR dimer with the asymptotic formulas. For
$\Gamma=0$, the baseline vanishes, while $D_{2}^{HSR}(t)$ turns
into Eq. \eqref{eq:D_A(t)}. The differences in the oscillating term
in Eq. \eqref{eq:D2_HSR(t)} and the diffusivity of the isolated dimer
in Eq. \eqref{eq:D_A(t)} are: (i) in the HSR model, the oscillations
are damped by the exponential factor, $\exp(-\Gamma t/2)$, (ii) the
amplitude depends on gamma, $\propto\sigma^{-1}(\Gamma^{2}+4J^{2})^{-1/4}$,
and (iii) there is an additional $\Gamma$-dependent phase shift $\arg(\Gamma-2i|J|)/2$.
Note that in a dimer, $D(t\rightarrow\infty)\rightarrow0$, thus,
the constant term in $D_{1}^{HSR}(t)$ is a consequence of the approximations
made (see SM for details \citep{SuppMat}).

On the short time scale, the height of the turnover point is practically
insensitive to the dephasing rate (for $0\leq\Gamma<|J|$), see Figs.
\ref{fig:FIG5}(a) and \ref{fig:FIG5}(b). In contrast, the first
minimum at $t=t_{m}\approx5\pi/(8|J|)\approx2/|J|$ becomes shallower
with increasing $\Gamma$, as there is enough time for the noise effect
to appear at this point. The position of the first minimum remains
approximately fixed (for $0\leq\Gamma<|J|$) allowing to approximate
$D_{A}^{HSR}(t=t_{m};\Gamma)\approx D_{A}^{HSR}(t=2/|J|;\Gamma)$.
In \ref{fig:FIG5}(c), we show the $\Gamma$-dependence of the first
minimum depth. In the considered examples, the dimer model slightly
overestimates the minimum depth.

While Eqs. \eqref{eq:D_A_HSR(t)}-\eqref{eq:D2_HSR(t)} are good approximations
for a dimer within the HSR model, they fail to capture the long-time
behavior in a chain {[}see Fig. \ref{fig:FIG5}(a){]}. Noise destroys
the localization ultimately resulting in classical-like diffusion
\citep{Moix2013,Chuang2016}, while in the dimer model the diffusivity
tends to zero. The mechanism behind the transition from the oscillatory
diffusivity to classical-like diffusion is the noise-induces hopping
between eigenstates. The mechanism can be understood by considering
the equations of motion for the density matrix expressed in the lattice's
eigenbasis (see SM \citep{SuppMat})
\begin{subequations}
\begin{align}
\dot{\tilde{\rho}}_{i,j} & =-i(\omega_{i,j}-i\Gamma)\tilde{\rho}_{i,j}+\Gamma\sum_{k,l}\kappa_{k,l}^{(i,j)}\tilde{\rho}_{k,l},\label{eq:HSR-eq-motion-eigenbasis}\\
\kappa_{k,l}^{(i,j)} & =\sum_{m}U_{i,m}U_{m,j}^{\dagger}U_{m,k}^{\dagger}U_{l,m},\label{eq:HSR-eq-motion-eigenbasis-kappa}
\end{align}
\end{subequations}
where $\tilde{\rho}$ is the reduced density matrix expressed in the
eigenbasis, $\omega_{i,j}$ is the difference of $i$-th and $j$-th
eigeneneregies, and $U$ is an orthogonal matrix with the real eigenvectors
arranged in rows. These equations shows that, in contrast to the closed
system, the populations of different eigenstates are coupled, allowing
the particle to spread beyond the initial region.

Equations \eqref{eq:HSR-eq-motion-eigenbasis} and \eqref{eq:HSR-eq-motion-eigenbasis-kappa}
can be further simplified by applying the often used secular approximation
\citep{NitzanBook2006} that captures the long-time dynamics. In this
approximation, the eigenstates' populations are decoupled from the
coherences, and obey (see SM for details \citep{SuppMat})
\begin{equation}
\begin{aligned}\dot{\tilde{\rho}}_{i,i} & =-\Gamma\tilde{\rho}_{i,i}+\Gamma\sum_{j}\kappa_{j,j}^{(i,i)}\tilde{\rho}_{j,j},\end{aligned}
\label{eq:secular-diagonal}
\end{equation}
where the coupling constants, $\kappa_{j,j}^{(i,i)}$ are determined
by the overlap of eigenstates' probability densities, $\kappa_{j,j}^{(i,i)}=\sum_{m}|U_{i,m}|^{2}|U_{j,m}|^{2}$.
The distributions of eigenstate widths in Fig. \ref{fig:FIG2}(a)
show that in the strong disorder regime all the eigestates are highly
localized and correspond mainly to dimers and monomers. Thus, Eq.
\eqref{eq:secular-diagonal} shows that in an open system the excitation
hops from the initially populated eigestates to adjacent spatially
overlapping eigenstates, and the hopping rate between eigenstates
$i$ and $j$ is $\Gamma\kappa_{j,j}^{(i,i)}$. The dashed line (HSR
appr.) in Fig. \ref{fig:FIG5}(a) shows the diffusivity obtained using
Eq. \eqref{eq:secular-diagonal}, which is in excellent agreement
with the full HSR model on the long time scale. \\

\emph{Conclusions.}--To summarize, this Letter systematically explored
the transient dynamics in a highly disordered 1D quantum lattice.
The scaling of the localization width with disorder allows us to identify
a transition between the weak and strong disorder regimes. In the
strong disorder regime, the lattice can be approximately partitioned
into monomers and dimers, thus establishing the random dimer model.
The resulting diffusivity exhibits unusual oscillations, with amplitude
decaying as $t^{-1/2}$ and inversely proportional to the disorder
strength. The first peak introduces a phase shift of $\pi/4$, with
the peak time and amplitude inversely proportional to the disorder
strength. In an open system, noise suppresses the oscillations on
short time scales, relaxes localization on intermediate scales, and
leads to constant diffusion for asymptotically large times. The mechanism
underlying the phenomenon is noise-induced hopping between lattice
eigenstates. 

The scaling predictions in this Letter not only help interpret recent
measurements but can also be quantitively tested on several platforms.
Recently, Anderson localization in 1D and 2D quantum lattices was
emulated by a fully controllable array of superconducting qubits \citep{Karamlou2022}.
The experimental results confirm the oscillatory behavior of the mean
squared displacement in the strong disorder regime and suggest the
possibility of experimentally verifying the quantitative predictions
reported in this Letter. Recent microscopy measurements of exciton
dynamics in disordered solids \citep{Akselrod2014,Delor2020,Blach2022,Blach2023}
also reveal oscillatory diffusivity, demonstrating the striking differences
between weak and strong disorder regimes. In follow-up papers, the
current study will be unified with a quantitative analysis of weak-intermediate
disorder \citep{CaoGroup} and is being extended to cavity polaritons
\citep{Georg2023,Qutubu}. Together, this and subsequent studies will
provide insights into optimizing transport properties for optoelectrical
and quantum information applications.\\

\begin{acknowledgments}
The work is supported by the NSF (Grants No. CHE1800301 and No. CHE2324300),
and the MIT Sloan fund, and is partly motivated by the joint experiment-theory
study in \citep{Blach2023}.
\end{acknowledgments}

\bibliography{bibliography}

\end{document}